\documentclass[10pt,conference,letterpaper]{IEEEtran}

\usepackage{setspace} 
\usepackage{enumerate}
\usepackage{xcolor}
\usepackage{cite}
\usepackage[cmex10]{amsmath}
\usepackage{amsthm}
\usepackage{graphicx}
\graphicspath{{./figs/}}
\usepackage[caption=false,font=footnotesize]{subfig}
\usepackage{bm}
\usepackage{pifont} 
\usepackage{mathrsfs}
\usepackage{amsfonts}
\usepackage{mathabx}

\usepackage{physics}

\usepackage{enumitem}

\usepackage{balance}

\newcommand{\nop}[1]{}

\usepackage{bibunits}
\defaultbibliography{main}
\defaultbibliographystyle{IEEEtran}

\begin{document}

\title{Topics in Quantum Networking}
\author{
\IEEEauthorblockN{Fengyou Sun}
\IEEEauthorblockA{
{Department of Information Security and Communication Technology}\\
{NTNU -- Norwegian University of Science and Technology}\\
Trondheim, Norway\\
sunfengyou@gmail.com
}
}

\maketitle


\begin{abstract}
It is an era full of imaginations and lack of impossibilities.
The knowledge boundaries have been being pushed back on and on.
The quantum age is on the edge of transforming quantum theories into quantum technologies.
We present a sketch of the advances of the quantum engineering towards quantum networks, based on which we discuss the research problems in performance analysis of quantum networks, with the goal to build a system theory for quantum network design and deployment.
\end{abstract}


\IEEEpeerreviewmaketitle

\section{Introduction}

In 1900, a new idea to describe nature was initiated by Max Planck, this is the quantum physics, which has had a profound impact in the following century \cite{kleppner2000one}\cite{zeilinger2000quantum}.
There are two entities in quantum physics \cite{kleppner2000one}, i.e., the quantum mechanics of matters that considers the physical phenomena at microscopic scale, and the quantum theory of fields that is concerned with the fields and interactions of the particles.
For example, the quantum mechanics can explain the order of elements in the periodic table \cite{scerri2019can} while the quantum field theory regards fermions and bosons as the two fundamental classes of particles and explains the identity of particles in the quantum mechanics \cite{kleppner2000one}.
The quantum physics is deemed to be the most successful theory with the most precise predictions in the history of science \cite{kleppner2000one}\cite{zeilinger2000quantum}.
Though experimentally tested, a consensus on the foundations and a satisfying interpretation elude the quantum theory.
There are some interpretations but they have mutual exclusions \cite{schlosshauer2013snapshot}\cite{cabello2017interpretations}, e.g., the wave functions collapse in the Copenhagen Interpretation but never collapse in the Many-Worlds Interpretation.
In order to understand the meaning behind the quantum calculation, a few physicists are on the way to reconstruct the quantum theory \cite{ball2013quantum}.
Specifically, the reconstructions are based on either the generalized theory of probability or information theory, particularly, the QBism theory avoids the paradoxes in the traditional quantum theory by relying on personal beliefs and expectations \cite{ball2017quantum}.
It is supposed that \cite{schlosshauer2013snapshot} the quantum reconstruction may provide  insights into the physical interpretation and quantum foundations, and lead to new theories, e.g., quantum gravity.

The quantum physics has an enormous implication on the technologies \cite{dowling2003quantum}. 
The first quantum revolution about wave-particle duality, which states that the matter particles can behave like light waves and vice versa, underpins the essential technologies in the modern society \cite{aspect2004john}, e.g., the electronics and photonics. 
Specifically, the transistors based on the electrical conduction are the basic elements of the integrated circuits, which are the core of the computing devices, and the laser are used to transmit information in the optical fibers that are the backbone of the classical Internet.
The manifestation of these information technologies marks the shift of our society from the Industrial Age to the Information Age \cite{kleppner2000one}\cite{aspect2004john}.
With the ability to control the individual quantum objects comes the second quantum revolution \cite{dowling2003quantum}, and humans are able to build artificial quantum systems with the unlikely prospects of the quantum cryptography, quantum teleportation, and quantum computation \cite{de2016quantum}, by utilizing the entanglement resources \cite{aspect2004john}.
The quantum information technologies have the potential to start up a Quantum Age \cite{aspect2016einstein}\cite{de2016quantum}.

The greatest challenge of the quantum technologies is to build a quantum computer \cite{dowling2003quantum}.
The quantum computers are proved through oracle separation to exist in a different computational complexity class from the classical computers \cite{tal2018oracle} and the quantum advantage in computation is confirmed in solving a linear algebra problem \cite{bravyi2018quantum}.
On the other hand, the quantum advantage in communication is experimentally demonstrated in a sampling matching problem \cite{kumar2018experimental}, i.e., the quantum communication allows to send exponentially fewer bits than classical communication for certain tasks \cite{bar2004exponential}\cite{buhrman2010nonlocality}.
Though the quantum computers are still pregnant in lab, the quantum networks are already under construction \cite{castelvecchi2018entangled}. Particularly, the quantum Internet \cite{kimble2008quantum}\cite{wehner2018quantum} is not only useful for connecting quantum computers together for distributed quantum computing but also has ``the potential to change the way in which people and organizations collaborate and compete, establishing trust while protecting privacy'' \cite{jason2017quantum}.

The remainder of this paper is structured as follows.
The implementations of the quantum bits, quantum computers, and the quantum networks are recapitulated in Sec. \ref{network}.
The research problems as regards the performance analysis of the quantum networks are discussed in Sec. \ref{performance}.
Finally, this paper is concluded in Sec. \ref{conclusion}.

\section{The Quantum Leaps}
\label{network}

\subsection{Quantum Bits}

The quantum bit or qubit, which was coined by Benjamin Schumacher in 1992, is the basic unit of information in quantum information theory \cite{tom2017birth}.
With respect to the physical implementation, a qubit is a physical system \cite{bennett2000quantum},
and the qubit state can be changed by changing the energy environment of the quantum system, which is defined as the Hamiltonian of the system \cite{NAP25196}.
There are many types of qubits each with its own strengths and weaknesses \cite{popkin2016quest}.
In general, the qubits based on the trapped ion are stable with slow operation, the superconducting qubits work fast with quick decoherence, the diamond qubits can operate at room temperature with the difficulty of entanglement generation, and the photonic qubits are useful for long-distance communication with probabilistic teleportation \cite{pirandola2015advances}.
However, these characteristics are not definite, for example, a new diamond defect is found to have both long coherence time and excellent optical properties \cite{rose2018observation}.

In mathematics, topology concerns the properties of spaces that are invariant under smooth continuous deformations \cite{asorey2016space}.
For half a century, topology has been broadly applied to physics, e.g., the topological matter \cite{castelvecchi2017strange}.
In topological matter, a collective particles can behave like an elementary particle, i.e., the quasiparticle, and the quantum states of multiple quasiparticles form the topological qubit.
The topological qubits are resilient to the outside interference due to the topological degeneracy of the quasiparticles that are neither bosons nor fermions but are anyons \cite{nayak2008non}.
Specifically, it can be achieved by either electron fractionalization or ground state degeneracy, i.e., storing the quantum information at two distinct places.
For example, the topological qubits are realized based on the fractional quantum Hall states \cite{stern2013topological} and the Majorana zero modes  \cite{kouwenhoven2018majorana}\cite{franz2018quantized}.
The topological qubits are coming into limelight with the potential for fault-tolerant quantum computing \cite{collins2006computing}\cite{nayak2008non}\cite{vishveshwara2011topological}\cite{NAP25196}.

In addition, an interesting realization of the qubits is based on the Schr\"{o}dinger's cat states \cite{leek2013storing}.
The cat state means the quantum superposition of the classically distinct states \cite{duan2019creating}, i.e., the Schr\"{o}dinger's cat is alive and dead at the same time \cite{haroche2013nobel}.
In an experiment \cite{wang2016schrodinger}, a wave packet of light composed of hundreds of particles is placed in two microwave cavities bridged by a superconducting artificial atom, which can be seen either as a cat state with two spatial modes or as an entangled pair of cat states, and it demonstrates that the Schr\"{o}dinger's cat lives or dies in the two cavities simultaneously, which is a manifestation of mesoscopic superposition and entanglement constructed from quasiclassical states and can be used to store the quantum information with redundancy \cite{wang2016schrodinger}\cite{osborne2016quantum}.
In another experiment, it shows that the Schr\"{o}dinger's cat states can be deterministically created by using a single trapped atom in a cavity to control the quantum states of the reflected light pulse \cite{hacker2019deterministic} and it is envisioned that the deterministic cat states can be useful for short- and mid-distance quantum communication due to the characteristics of loss correction \cite{bergmann2016quantum}.

The qubit is the basic element in quantum information processing \cite{stajic2013future}, which encompasses quantum computation, quantum cryptography, quantum communication, quantum simulation, and so on.
For sake of stability, fault tolerance, and scalability, quantum error correction \cite{NAP25196} is proposed based on the logical qubits, each of which is emulated by a number of physical qubits.
It is depicted that the development of quantum information processing includes seven stages \cite{devoret2013superconducting}, from the operation on physical qubits through the operation on logical qubits to the fault-tolerant quantum computing.

\subsection{Quantum Networks}

The quantum Internet \cite{kimble2008quantum} is envisioned to enhance the classical Internet by enabling quantum communication among arbitrary network users \cite{wehner2018quantum}.
There are three elemental quantum hardware components that make up the quantum Internet \cite{wehner2018quantum}, i.e., the quantum channel, quantum repeater, and end nodes.
Specifically, the quantum channel is the physical medium to transmit the qubits, the quantum repeater is used to enable the quantum networking of arbitrary distances, and the end nodes are used for information processing, e.g., quantum computers.
The quantum repeater establishes the connectivity through entanglement swapping \cite{munro2015inside} and then the information is transmitted through quantum teleportation \cite{zeilinger2018quantum}, which is envisioned to be the quantum feature of the future Internet  \cite{zeilinger2018quantum}.
A review of quantum teleportation is available in \cite{pirandola2015advances}, e.g., the deterministic and probabilistic teleportation, the active and passive teleportation. 
In addition, the quantum networks need the quantum memory \cite{simon2010quantum} for the storage and process of the quantum information \cite{pirandola2016unite}.
It is depicted that there will be six implementation stages of quantum Internet \cite{wehner2018quantum}, ranging from the trusted repeater networks, which is the current implementation status, 
through the quantum memory networks, 
to the ultimate quantum computing networks.

The optical light has the advantage of long-distance communication \cite{pirandola2016unite} and photonic channels are usually used to establish the quantum links between the quantum repeaters and between the end nodes \cite{wehner2018quantum}, e.g., the free-space channels and the fiber-based channels.
The quantum information can be encoded by two types of observables \cite{andersen2015hybrid}, i.e., the discrete variables, e.g., the polarization state, and the continuous variables, e.g., the intensity and phase of the electric field of the electromagnetic wave.
Contrast to the light-based quantum communication, the quantum memory and the quantum computer are usually matter based, e.g., the solid-state quantum memories, and a quantum interface is required to convert the light-based quantum states to the matter-based quantum states and vice versa \cite{kimble2008quantum}\cite{pirandola2016unite}.
It is proposed that the hybrid approaches are needed to combine the features of both the discrete-variable and the continuous-variable technologies and to integrate the light-based communication and the matter-based storage and processing for a scalable quantum Internet \cite{pirandola2016unite}.
In addition, the integration of the quantum networks and the classical networks is discussed in \cite{sasaki2017quantum}.

The quantum computers, which are not mandatory for some quantum network protocols \cite{wehner2018quantum},
are the essential components of the quantum computing networks, where the quantum computers can arbitrarily exchange information through the quantum network.
On the other hand, it is an intriguing way to build the quantum computers by connecting a few quantum systems together \cite{monroe2016quantum}.

\subsubsection{Quantum Computers}

There are three types of quantum computers \cite{NAP25196}, i.e., the analog quantum computers, the digital noisy intermediate-scale quantum computers, and the fully error-corrected quantum computers, where the first type of quantum computers are based on the analog methods, e.g., quantum annealing and quantum simulation, and the last two types of quantum computers are based on quantum logic gates.
Since the analog quantum computers can be built by controlling the Hamiltonian of the quantum system in a straightforward way without full error-correction, it has an advantage with respect to the digital quantum computers to solve practical problems in the near term and is deemed to become obsolete due to the control difficulty and be surpassed by the digital quantum computers in the long term \cite{preskill2018quantum}\cite{NAP25196}.
The practical utility, quantum supremacy, and fault-tolerant computing can be seen as the milestones of the quantum computer development, and the accumulation of the physical and logical qubits in the quantum systems can serve as the basis to measure the development progress \cite{NAP25196}, e.g., the quantum volume \cite{cross2018validating}.

It is anticipated that the future quantum computer would have a hybrid architecture \cite{popkin2016quest}, with the superconducting qubits running algorithms, the trapped ion qubits forming memory, and the photonic qubits communicating signals.
The topological methods have the advantage of achieving the logical qubits with far less physical qubits \cite{NAP25196}.
In addition, the classical computers are needed to control the quantum operations and to implement the computations for the quantum error correction \cite{van2019electronic}\cite{NAP25196}.
To enable transparent application development and network management, the quantum networks need the quantum software to allow the quantum algorithms and quantum protocols to connect to the quantum hardware \cite{zeng2017first}\cite{mueck2017quantum}\cite{wehner2018quantum}.
The quantum software are layers of software tools, for example, the quantum programming languages and compilers for quantum application development \cite{chong2017programming}, and the quantum error-correcting code for fault tolerant quantum computation \cite{campbell2017roads}.
It is depicted that the quantum algorithms will feature hybrid approaches \cite{zeng2017first} to combine both the classical and the quantum processors for the noisy intermediate-scale quantum computers \cite{preskill2018quantum}, which are different from the quantum algorithms for the noiseless and large-scale quantum computers \cite{montanaro2016quantum}, and the quantum network software stack will be a synergy of the quantum computing stack and the classical network stack \cite{wehner2018quantum}.

Albeit the passion in quantum computer \cite{gibney2014physics}\cite{lekitsch2017blueprint}\cite{gibney2017physicists}, there are pessimistic voices against it, arguing that a quantum computer needs to control an astronomical amount of continuous parameters with high precision that is impossible \cite{dyqkonov2019when} and that the quantum systems are inherently noisy \cite{kalai2016quantum}, and there are no definite answers on when the useful quantum computers appear \cite{NAP25196}\cite{schneider2019quantum}\cite{maslov2019outlook}.

\section{Opportunities in Performance Analysis}
\label{performance}

Technically, the development of quantum Internet requires a hybrid of technologies, which combine the features of both discrete variable systems and continuous variable systems \cite{pirandola2016unite}.
Theoretically, it is necessary to formulate a system theory \cite{sun2018performance} for the dimension of the network dynamics, i.e., backlog, delay, and throughput, 
to deal with the quality-of-service requirements of the network applications 
and to help design and deploy the quantum networks.

\subsection{Beyond the Analysis of Classical Networks}

The classical methodology for network performance analysis is queueing theory, which was initiated for teletraffic analysis in telecommunication networks by Agner Krarup Erlang
 in 1909 \cite{erlang1909theory} and was revived to analyze computer networks by Leonard Kleinrock in the 1960s \cite{kleinrock2002creating}.
We regard the queueing analysis as the canonical approach because of its pervasiveness and irreplaceability in networking.
The queueing theory deals with the accumulation of the interarrival time and service time in time unit \cite{kleinrock19756queueing} or the accumulation of arrival quantity and service quantity in bit unit \cite{sun2018performance}, based on which the performance measures are defined, e.g., backlog as the accumulated workload in the queue or the whole system, delay as the waiting time or sojourn time with respect to different scheduling schemes, and throughput as the traffic amount passing through the queueing system.
A communication network is seen as a network of queues, due to the irregular randomness of the queue output, it is difficult to analyze the queueing networks and the queueing analysis based on the accumulation of bits usually has an advantage over the analysis based on the accumulation of time.

Since the system theory is a mathematical tool to facilitate the network design and deployment in real world, the fundamental challenge of formulating a system theory lies in what form the network is and will be of, or what form of network the theory indicates to build with a better performance in terms of some performance measures.
If we build the quantum internet as an alternative or augmentation to the classical internet architecture, i.e., the quantum internet serves to connect the quantum computing devices here and there for communication, which is exactly the initial motivation of the classical internet \cite{leiner2009brief}, a few research issues should be addressed, considering the quantum network uniqueness.

\subsection{Quantum Network Characteristics}

The quantum Internet is not only a hybrid architecture in terms of the diverse implementations, but also a hybrid system by the complex nature of the quantum physics.

\begin{enumerate}[wide, labelwidth=!, labelindent=0pt]
\item
\emph{Quantum diversity.} 
Compared to the classical communication, the quantum communication has additional physical resources, i.e., the superposition and entanglement, and the quantum concepts usually have a manifold characteristic.
From the information-theoretic perspective, the quantum channel capacity has many concepts \cite{bennett2000quantum}\cite{bennett2014quantum}, i.e., the classical capacity, quantum capacity, private capacity, entanglement-assisted capacity, etc.
In addition, it shows that \cite{ebler2018enhanced} the ability to combine quantum channels in a superposition of orders can boost the rate of communication beyond the limits of conventional quantum Shannon theory, which is due to the quantum causality with indefinite causal order \cite{brukner2014quantum}\cite{rubino2017experimental}\cite{ball2017howquantum}.
These different capacity concepts describe different facets or capabilities of the quantum internet, which means that the quantum internet can be used for different purposes with different protocols or technology supports, e.g., the transformation of classical information, quantum information, or private information.
On the other hand, the information can be either classical bits or quantum bits, for example, the information coming from the classical source can be classical, the information coming from the quantum source can be quantum.
If such information is to be transmitted through the quantum internet, through quantum states, the classical information should be coded into the quantum states, while the quantum information may need to be converted from one type of quantum system state to another quantum system state for transmission, e.g., the light-matter transformation \cite{kimble2008quantum}\cite{pirandola2016unite}. 
The diversity of the quantum concepts implies the diversity of the quantum networks, e.g., the networking analysis should consider the multiplexing of the classical information and the quantum information in a heterogeneous network of diverse quantum channels, e.g., the communication links and storage.

\item
\emph{Information additivity and causality.}
The performance analysis considers the information quantity, i.e., the amount of storage space \cite{horodecki2005partial}, rather than the information content.
The information transmission follows the causality principle \cite{pawlowski2009information} and the transmitted information quantity is additive. 
However, the existing formulas of quantum channel capacity are non-additive in general, which indicates an incomplete understanding of the quantum channels and poses a challenge to the network service modeling in performance analysis.
Particularly, it is interesting to study the probabilistic characterization of the quantum information quantity, e.g., whether it is a concept in the classical regime or in the quantum regime \cite{zurek2006decoherence}.
In addition, it is interesting to muse on the performance analysis of the hybrid of the quantum network and the classical networks and it is interesting to consider the quantum network coding \cite{hayashi2007quantum}\cite{leung2010quantum}.
Besides the quantum hardware, there is also a stack of quantum software. The diversity of the software and the protocols also imply the diversity of the analysis.
Moreover, the quantum signal processing techniques also have an impact on the quantum network performance analysis, e.g., quantum signal modulation, quantum source and channel coding.

\item
\emph{Metric uncertainty.}
Analogical to the classical network performance measures, it is straightforward to define the quantum network performance measures as the backlog, delay, and throughput.
However, it is unknown whether these measures are enough or not to evaluate the quantum networks, whether to integrate the quantum mechanics of entanglement and superposition into these existing measures or to define new measures.
The metric system should evolve along with the quantum network evolution from the prototype to the commercialization. 
In practice, the potential of quantum networking lies in that it is unknown how many benefits the quantum internet will bring to the society besides the analogical functionality of the classical internet, in other words, it is unknown what the transformative feature will be due to the quantum coherence and correlation.
This indicates opportunities of new applications and technologies.
On the other hand, there will be new regulations on network operation and management, which influences not only the network architecture but also the network model and analysis, and indicates an extra challenge to quantum network performance analysis.
\end{enumerate}

\section{Conclusion}
\label{conclusion}

The quantum physics is a growing theory still without a definite answer to the nature of reality and the quantum engineering is an expanding field with more and more advanced quantum technologies.
No mater how much the quantum Internet mimics the classical Internet in terms of the architectures and applications at present, it is hard to deny the possibility that the future quantum Internet will be fundamentally different from the present design.
Considering the diverse realizations of the quantum bits, quantum devices, quantum networks, and quantum software, we envision an inclusive framework for the performance analysis of quantum networks with a varied choice of methodologies to address the different implementations.

\balance
\bibliography{main}
\bibliographystyle{IEEEtran}

\end{document}